\documentclass[twocolumn,showpacs,prl]{revtex4}

\usepackage{graphicx}
\usepackage{amssymb}

\begin{document}

\preprint{APS/123-QED}

\title{Very Long Time Magnetization Decays in Spin Glasses}

\author{G.~G.~Kenning}

\author{G.~F.~Rodriguez}
\affiliation{Department of Physics, University of California\\
Riverside, California 92521-0101}
\author{R.~Orbach}

\affiliation{Department of Energy, Office of the Director of Science,\\
Washington DC\\}

\date{\today}

\begin{abstract}
It is currently believed that the decay of the thermoremanent
magnetization in spin glasses is composed of two terms;  The
"stationary" term which does not depend on the sample history and
dominates the short time decay ($<1s$) and a long time aging term
which depends on the samples history. The sample history includes
both the thermal history and the time spent in a magnetic field
(waiting time) before that field is removed. We report finding a
third component of the decay at times much larger than the waiting
time. This decay is independent of the waiting time, but part of
the same mechanism that produces aging.  We explain this decay in
terms of the initial state distribution that is developed during
the cooling process.  This explanation is tested by performing
Iso-thermoremanent magnetization decay measurements. Follow up
calculations using the Spin Glass phase space barrier model
suggest that the long time decay is logarithmic and that the
maximum aging time is approximately three thousand years, in this
sample, at a measuring temperature of .83 $T_{g}$.
\end{abstract}

\maketitle

Understanding the effects of disorder on states of matter has been
the subject of intense study for almost 100 years.  In the early
1970s the realization that highly disordered magnetic materials
with random interactions undergo an apparent phase
transition\cite{Can72, EA75}, sparked considerable interest and
excitement in the condensed matter community.  This discovery of
the so-called spin glass phase was marked by a large experimental
effort by many researchers in an attempt to try to understand the
physical parameters of the underlying phase space\cite{Bin86}. One
very interesting property of the spin glass phase was the
observation that there exists a large distribution of relaxation
times extending from atomic fluctuation time scales through what
are believed to be geological time scales.

In 1983 it was shown that materials in the spin glass phase have
memory effects on time scales ranging from milliseconds, to
seconds, to days and beyond\cite{Cham83}.  To date much
information has been obtained about the structure of the spin
glass phase space from the study of aging effects. The classic
measurement of these memory effects is the Thermoremnant (TRM) or
complementary Zero Field Cooled (ZFC) magnetization decay
measurements. In the TRM experiments the sample is cooled, in a
small constant magnetic field, through its transition temperature,
to a measuring temperature $T_{m}$. After waiting a time $t_{w}$,
in the magnetic field, the field is rapidly removed and the
consequent magnetization decay of the sample measured. The decay
is strongly dependent on the waiting time ($t_{w}$). This is
called aging. It has recently been observed\cite{Rod03} that the
magnetization decays scale with $t_{w}$ for a spin glass cooled,
sufficiently rapidly, to a particular measuring temperature.

While aging has been the primary focus of TRM measurements, it is
not the only contribution to the decay.  Upon cooling the spin
glass sample from a high temperature, in a magnetic field, the
magnetization generally follows a Curie or Curie Weiss like
behavior. Cooling the spin glass further, through its transition
temperature, the magnetization appears to remain approximately
constant, at a value $M_{fc}$. This is an indication that the
spins freeze in approximately random directions with a net bias
due to the field.  Upon removal of the field there is a rapid
decrease in the sample magnetization, from a value of $M_{fc}$,
before entering the aging regime. Various measurements including
AC susceptibility\cite{Tho81, Ref88}, muon decay\cite{Mac83},
neutron spin echo\cite{Mez79} and NMR\cite{All85} each provide
some evidence that this rapid decay is power law in nature.  The
power law term is independent of the waiting time and has been
called the stationary term (i.e. stationary with respect to the
waiting time).

Current belief \cite{Bou95,Cou93, Bou98} is that the full
magnetization decay can be described by  a power law plus an aging
term

\begin{eqnarray}
M_{TRM} =  At^{-\alpha}+ M(\frac{t}{t_{w}})  \label{eq:one}
\end{eqnarray}

In this study we show that there exists another part of the decay
which is independent of the waiting time and occurs in the time
regime $t>>t_{w}$.  This is a new region of the decay and is not
related to previous discussions of $t>>t_{w}$\cite{Alba86,Rie93}.
Analysis of this long time regime suggests that it is independent
of the waiting time and is not a continuation of the initial
stationary (power law) term.

All measurements in this work were performed on an alloy
$Cu_{.94}Mn_{.06}$ that has a transition temperature $T_{g} =
31.5K$. The sample has been the focus of many previous studies and
as such has been well-characterized\cite{Ken91}.  Measurements
were performed on a home built DC SQUID magnetometer that allowed
for good control of the rapid temperature quench and subsequent
waiting time. All measurements in this study were performed at $T=
26 K$ which corresponds to reduced temperature of $0.83 T_{g}$ .

\begin{figure}
\resizebox{\columnwidth}{4.0in}{\includegraphics{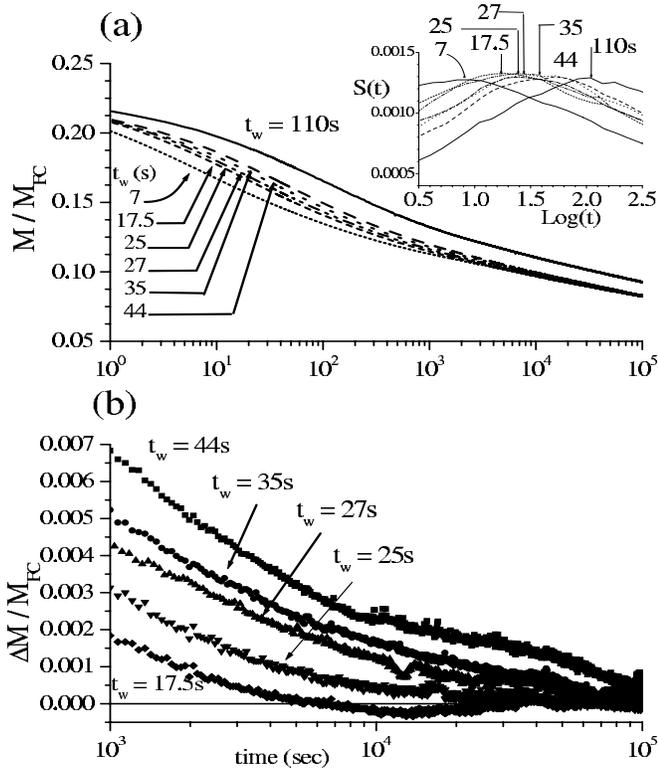}}
\caption{\label{fig:Cool1} A(a) Decay curves for effective waiting
times of: 7,17.5, 25, 27, 35, 44 seconds. In the inset, the S(t)'s
for each curve used to determine the effective waiting time. (b)
the subtraction of the fastest decay from the other curves,
$\Delta M~=~M(t_{w}=7s)- M(t_{w})$,for $t_{w} =
17.5(\blacklozenge),25(\blacktriangledown),27(\blacktriangle),35(\bullet),44(\blacksquare)$.
It can be seen that the M($t_{w}$=17) curve begins to overlap
M($t_{w}=7$) at approximately 5000s.}
\end{figure}

We report very long time TRM decay measurements ($10^{5}$ seconds)
for very short effective waiting times ( $7s <t_{w}<110s$ ). The
data is plotted in Figure $\ref{fig:Cool1}$a. The decays were
obtained using identical fast cooling protocols with a small
additional waiting time added before the magnetic field is shut
off.  It is this small additional waiting time that changes the
age of the curves in Figure 1a. The ages of the individual curves
were determined from the peak in the relaxation curve
($S(t)=-\frac{1}{H} dM/dln(t)$) plot shown in the inset of Figure
$\ref{fig:Cool1}$a. It can be observed that as the effective
waiting time increases the aging portion of the curves separate.
This is a well-known effect of aging\cite{Sitges}.  We observe for
the short waiting times curves that the decays overlap each other
at very long times. Since all of the decay curves either follow or
approach this long time decay, we conclude that it is independent
of the waiting time and hence distinct from the aging regime. We
plot the difference between the 7s waiting time curve and the
longer waiting time curves in Figure $\ref{fig:Cool1}$b. The range
over which this difference goes to zero determines the onset of
the third component of the decay for each of the waiting times. It
can be observed that, within error, the decay of the 17s waiting
time decay begins overlapping the 7s waiting time decay at
approximately 5,000 seconds.   It can also be observed that the
25s waiting time decay overlaps, within error, at approximately
35,000 seconds while the 27s waiting time decay overlaps at
approximately 75,000 seconds. The 35s and 45s waiting time decays
systematically approach the 7s waiting time decay, but the
magnetization difference does not go to zero within a measuring
time of 100,000 seconds.  This analysis is subject to the errors
inherent within the measuring system. Sources of error include a
high frequency point-to-point fluctuation and low-frequency
"wobbles" (possibly due to small temperature fluctuations
$<10mK$). The signal to noise ratio, at 100,000 seconds, is
approximately 100 to 1.

\begin{figure} [ht]
\resizebox{\columnwidth}{4.0in}{\includegraphics{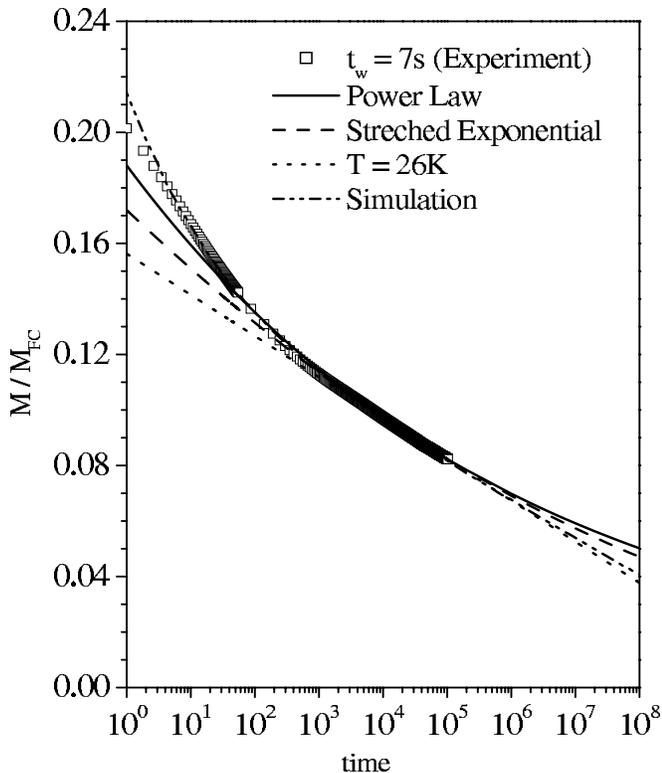}}
\caption{\label{fig:Fits} Three fits to TRM($t_{w} = 7$s),fitted
to $t >$ 5000s. The three fitting functions are: power law,
stretched exponential and logarithmic. For power law
($M=M_{0}t^{-\alpha}$), $M_{0} = 0.1881$ and $\alpha = -0.072$.
For stretched exponential ($M=M_{0}$exp$[t^{-\alpha}/\tau_{0}]$,
$M_{0} = 1.48, \tau_{0} = 10^{-13} $ and $ \alpha = -0.02559$. For
logarithmic fit( $M = M_{0} + Log(t^\alpha)$), $M_{0} = 0.156$ and
$\alpha = -0.0148$.}
\end{figure}

It therefore appears that the aging effect is finite in time
extent and at long enough time scale gives way to a long time
component of the magnetization decay that is independent of the
waiting time.  Unfortunately the region of the overlap
(5,000-100,000s) between the 7s and 17s waiting time decays gives
only one and a half decades of measurement time with which to fit
a function, precluding an absolute determination of the form of
the fitting function.  We find that we can fit this range to a
power law, a stretched exponential and a logarithmic
function(Figure $\ref{fig:Fits}$). Values of the fitting
coefficients and exponents are given in the figure caption.  We
take a data point every 3 seconds for a total of approximately
20,000 measurements in the overlap range. The large number of
measured points and small signal to noise ratio give a very small
value for the $\chi^2$ of the fitting functions. The stretched
exponential is the worst fit with $\chi^2=$5.8x$10^{-7}$. The
power law fit gives $\chi^2=$7.0x$10^{-8}$,  while the logarithm
gives the best fit with $\chi^2=$1.2x$10^{-8}$,

It is possible that Eq.$\ref{eq:one}$ is the correct description
of the decay and that we are observing at long times, the short
time stationary term after the aging decay has ended. To test this
we have determined the upper limit of the power law exponent for
the sample, measured at.83 $T_{g}$, using a TRM magnetic field of
20G and a waiting time of 10,000 seconds (data is displayed
in\cite{Rod03}). The short time decay in the TRM should be
dominated by the power law decay of the short time stationary
term, Eq.$\ref{eq:one}$. Fitting over the first 3 seconds of this
TRM decay we have determined that the upper limit for the exponent
$\alpha$ is 0.01. This compares with an exponent of the power law
$\alpha~=~0.072$ found for the very long time decay. This value is
also much larger than the value reported from ac susceptibility
experiments\cite{Sitges} for the short times stationary term. This
significant difference suggests that the short time and long time
decays are of different origin.

We now address the question as to whether this long time decay is
an additive term. Subtraction of any of the fitting functions from
the general decay curves produce sets of curves that no longer
scale as t/$t_{w}$. In fact, we are not able to scale the
subtracted data using any of the standard scaling
techniques\cite{scaling}. Unless we are to abandon the concept
that the aging decays as t/$t_{w}$, or that the decays scale at
all, we are led to the conclusion that this long time decay is not
simply an additive term like the short time stationary decay. This
analysis implies that this long time stationary decay is
intrinsically related to the aging process and does not occur
until the effects of aging are effectively over. An understanding
of both, the mechanisms responsible for, and form of, this
long-term decay can be found through analysis of the barrier
model.

The Barrier Model of the spin glass phase space was developed to
explain aging in spin glasses.  Inspired by the Parisi\cite{Par80}
solution of the first principles Sherrington
Kirkpatrick\cite{She78} model of spin glasses and grounded in
experimental analysis, the barrier model implies that spin glass
phase space can be described by diffusion of the system within a
phase space, defined by a set of energy barriers which grow
linearly as a function of Hamming distance\cite{Led91}. Generally
a minimum barrier $\Delta_0$ is defined and the height of  barrier
$N$ is equal to the product of $N$ and $\Delta_0$. The model
simplifies the hierarchically ordered tree structure of the Parisi
solution by collapsing local branches of the tree(states) into a
single state with the appropriate degeneracy multiplication
factor(this corresponds to the product of the barrier number N and
the branching ratio r). Diffusion within the Barrier Model is
essentially a trade-off between the geometrically growing number
of states as a function of increasing barriers and an
exponentially decreasing probability for Boltzman hopping over the
increasing size of the barriers. This model has proven to be very
successful in quantitatively modelling aging decays, including
temperature cycling. \cite{Joh96}

\begin{figure} [h]
\resizebox{\columnwidth}{4.0in} {\includegraphics{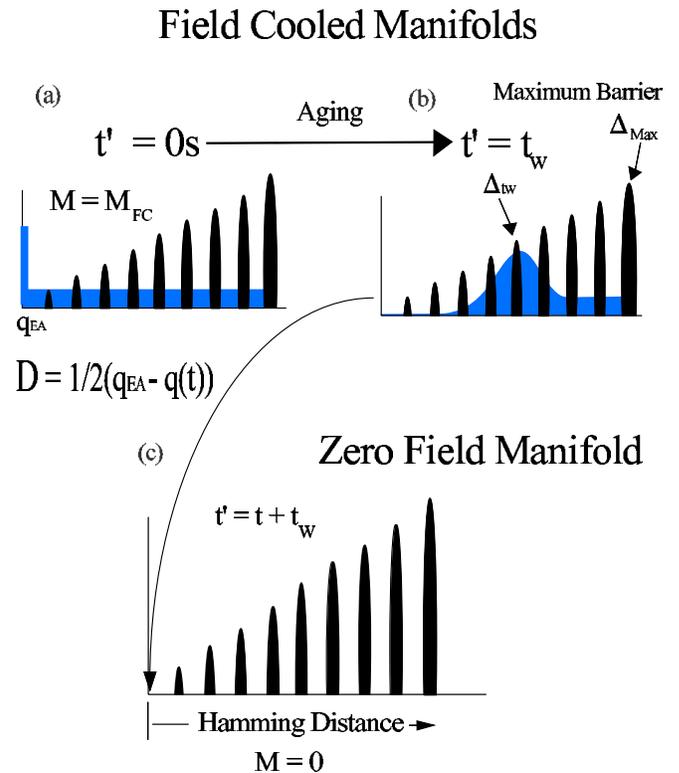}}
\caption{\label{fig:BarModel} TRM experiment interpreted using a
schematic of the Barrier model.  During the cooling process, there
is a "seeding" of the phase space. An initial distribution is set
up.  During the waiting time, states begin to diffuse into the
phase space.  The barrier associated with waiting time is
$\Delta_{tw}$.  The highest barrier is $\Delta_{max}$}
\end{figure}

The initial state occupation is formed upon cooling the spin glass
through its transition temperature to some measuring temperature,
in a small magnetic field (Figure $\ref{fig:BarModel}$). Magnetic
field cycling during the waiting time has previously been
employed\cite{Ken95}  to probe the redistribution of states in
going from the field cooled manifold to the zero field manifold
and back.The initial state occupation was determined to be a delta
function located at zero energy barrier with a possible background
contribution from states close to the delta function. In reality,
this study probed the initial distribution within the zero field
manifold (after the field was cut) and this distribution was also
taken to be the initial distribution in the field cooled manifold
after initial cooling. In the present study we find that it is not
the case that the initial distribution in the field cooled
manifold is a delta function at the origin.  Instead, we are only
able to fit the longtime decay data by assuming that the states
formed during the cooling process are distributed uniformly over
all possible states (i.e. over all barriers). This has several
significant implications. First, it has a self similar background
which is necessary for understanding the temperature dependence of
the TRM. Secondly, it suggests that correlations of all possible
strengths are formed during the cooling process. Finally, the
uniform background produces a logarithmic decay, implying a
limiting barrier height and hence a limit on aging.

In Figure $\ref{fig:Fits}$ we plot a fit of the whole 7s waiting
time TRM decay using the results of the barrier model
($\Delta_{0}=.12$, $r=1.048$ and $N_{max}=800$ barriers), evolved
from a constant background initial state, coupled with a power law
term to model the "short time" stationary part of the decay (Eq.
1).  Values for the power law were those determined from the
10,000 second waiting time data, previously mentioned.
Extrapolating (straight line logarithmic dependence of the barrier
model) to zero magnetization, we find a limiting age
(corresponding to a maximum barrier) of $t_{max} \approx
$1.1x$10^{11}$ seconds. This corresponds to approximately 3000
years.  This remarkable result implies that the phase space, at
least in this sample, is limited in the time domain.

\begin{figure} [ht]
\resizebox{\columnwidth}{4.0in}
{\includegraphics{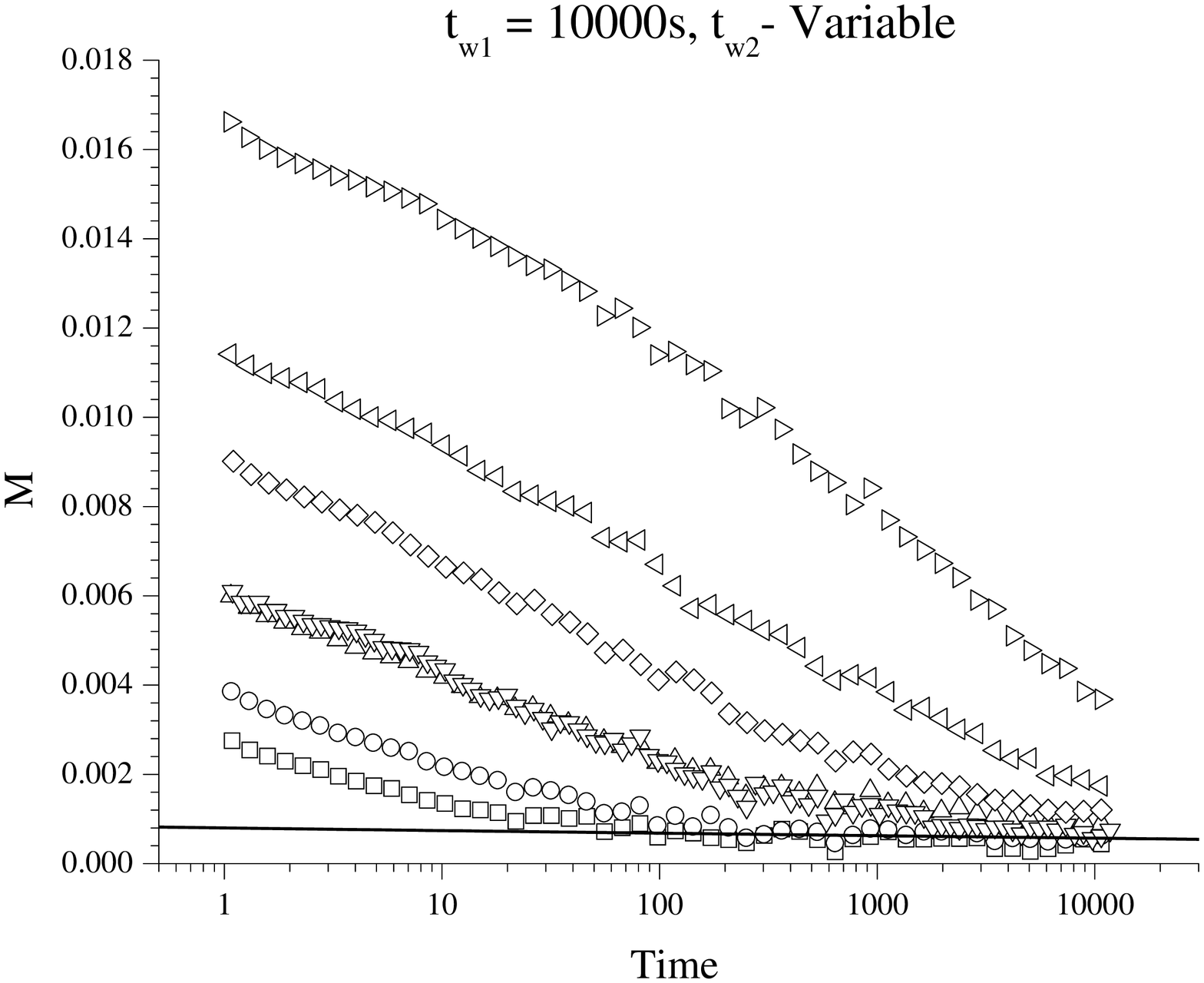}} \caption{\label{fig:IRM}
IRM decays. We use $t_{w1}$ = 10000s and vary $t_{w2}:
10(\square), 25(\bigcirc), 100(\vartriangle), 150(\triangledown),
300(\lozenge), 1000(\vartriangleleft)$ and $
3000(\vartriangleright)$ seconds. The fitted curve is a power law
with the power, $\alpha = -0.01$.}
\end{figure}

We can test the assumption of an initial uniform state
distribution over all barriers by performing an Isothermal Remnant
Magnetization (IRM) measurement.  In this experiment the sample
was cooled in zero magnetic field, producing an initial
distribution of states within the zero field (zero magnetization)
manifold.  The sample is kept in zero field for a time $t_{w1} =
10,000$ seconds. A small magnetic field (20G) is then applied for
a time $t_{w2}$. After $t_{w2}$ the magnetic field is removed and
the decay observed. The data is plotted for several different
values of $t_{w2}$ in Figure \ref{fig:IRM}.

We start our analysis of the data with the assumption that the
field cool (FC) and zero field (ZF) manifolds are equivalent. This
assumption has strong experimental backing.\cite{Lun85} The
cooling procedure in the IRM should therefore produce an
equivalent distribution of states in ZF manifold as the TRM
produces in the FC manifold. Waiting for a long time $t_{w1}$
allows enough time for states, which contribute to aging, to
diffuse over significantly large barriers in the ZF manifold. When
the magnetic field is applied a small number of these aging states
diffuse back to the zero barrier and jump to the FC manifold.
These states then begin to age, starting from the zero energy
barrier, in the FC manifold for a time $t_{w2}$. Removal of the
magnetic field then produces a measurable decay of the states back
from the FC manifold to the ZF manifold.  If our assumption that
the background is due to the cooling effect, then the uniform
background should be located deep within the ZF manifold and
should not be observed in the IRM experiment. This is exactly what
is observed. We do find an aging decay followed by a very small
power law decay with an exponent compatible with that observed in
the short times stationary measurements.

In conclusion, we have measured very long time TRM measurements
with rapid cooling protocols and small waiting times.  We find
that that the aging time is finite in extent and that there is a
longtime decay which is independent of the waiting time.  This
decay is different from the short times stationary term and
related to the mechanisms responsible for producing aging in spin
glasses.

\end{document}